# Contact-induced doping in aluminum-contacted molybdenum disulfide

**Yoshihiro Shimazu\*, Kensuke Arai, and Tatsuya Iwabuchi**

Department of Physics, Yokohama National University, Yokohama 240-8501, Japan

\*E-mail: yshimazu@ynu.ac.jp

**Abstract**

The interface between two-dimensional semiconductors and metal contacts is an important topic of research of nanoelectronic devices based on two-dimensional semiconducting materials such as molybdenum disulfide ($MoS_2$). We report transport properties of thin $MoS_2$ flakes in a field-effect transistor geometry with Ti/Au and Al contacts. In contrast to widely used Ti/Au contacts, the conductance of flakes with Al contacts exhibits a smaller gate-voltage dependence, which is consistent with a substantial electron doping effect of the Al contacts. The temperature dependence of two-terminal conductance for the Al contacts is also considerably smaller than for the Ti/Au contacts, in which thermionic emission and thermally assisted tunneling play a dominant role. This result is explained in terms of the assumption that the carrier injection mechanism at an Al contact is dominated by tunneling that is not thermally activated.

## 1. Introduction

Molybdenum disulfide (MoS$_2$) is one of the promising layered semiconducting materials that can be used to realize novel nanoelectronic and nano-optoelectronic devices.[1,2] Its ultrathin structure, with a monolayer of thickness < 1 nm, enables scaling of devices, which is not possible using conventional semiconductors such as Si. Although it has been well known that the transport properties at the interface between MoS$_2$ and contact metals significantly affect the electrical characteristics of the devices,[3-10] understanding of this issue is still in its infancy. In past studies, Au is among the most widely used contact metals. A thin sticking layer of Ti or Cr is often inserted between Au and substrates. For a Au contact, ohmic characteristics are commonly observed at room temperature. This does not imply the absence of a Schottky barrier at the contact.[11] The main carrier injection mechanisms at the Schottky barrier for the Au contact are known to be thermionic emission and thermally assisted tunneling of electrons.[3,7,11] Therefore, with decreasing temperature, the interface resistance increases exponentially. If the Schottky barrier is eliminated by some technique, such strong temperature dependence could be suppressed, and low contact resistance would be achieved. Although many studies have focused on reduction or elimination of the Schottky barrier,[12-15] convenient and reliable method for this purpose has not been made available. Doping of MoS$_2$ is another important topic of research. Conventional doping methods for Si[16] cannot be applied for MoS$_2$ because it is atomically thin. Doping of MoS$_2$ using a metallic contact would be extremely useful. The contact-induced doping should not be accompanied by MoS$_2$ degradation, which may be caused by chemical doping, such as adsorption of dopant molecules on the surface of MoS$_2$.[13,14,17-19] Contact-induced doping in MoS$_2$ has rarely been reported. Although hole doping by palladium contacts has been reported,[9] its mechanism is not clear.

In this paper, we report the transport properties of MoS$_2$ flakes with Al and Ti/Au contacts, and compare the experimental results between the different contacts. There are only limited reports on the transport properties of transition metal dichalcogenides contacted by Al.[6,8,20] In our experiment, the Al contact exhibited considerable electron doping effect, which led to high carrier density in MoS$_2$ accompanied by a substantial shift of threshold gate voltage in transfer characteristics. Furthermore, temperature dependence of the two-terminal conductance indicates that the carrier injection mechanism at the Al contact is governed by electron tunneling, which does not involve thermal activation. This was not observed in previous studies that utilized various contact metals and Al/Au contact. We found that depositing a top Ti/Au layer on the Al contact significantly altered the transport properties.

## 2. Experimental methods

We exfoliated thin MoS$_2$ flakes from a bulk crystal (SPI Supplies), and deposited them on a Si substrate with a 270-nm-thick SiO$_2$ layer, using adhesive tape and a gel sheet.[21,22] The source and drain electrodes were fabricated using photolithography and electron beam deposition of metals. The devices were annealed at ~110 °C for up to 15 h in vacuum (~3×10$^{-4}$ Pa). The highly n-doped Si substrate was used as a back gate. The thicknesses of the MoS$_2$ flakes were measured using an atomic force microscope. We compared the transport properties of the devices with thin flakes of thicknesses ~15 nm each. For transport measurement at varying temperatures, the devices were contained in a sample cell of a cryostat filled with helium gas for heat exchange. Drain-to-source current $I_d$ was measured as a function of drain-to-source voltage $V_d$ and gate voltage $V_g$. Representative data taken from several devices with Ti (15 nm)/Au (75 nm) or Al (90 nm) contacts are presented below. Except the deposited metals, fabrication processes were the same for all the samples that we analyzed in the present study.

## 3. Results and discussion
### 3.1 Transport properties at room temperature

In Figs. 1(a)–1(d), the output characteristics ($I_d$–$V_d$ curves) and transfer characteristics ($I_d$–$V_g$ curves) at room temperature are compared between the MoS$_2$ flakes contacted by the Ti/Au and Al electrodes. In the case of the Ti/Au contact, n-channel FET characteristics are clearly observed, as reported in many literatures; below the threshold voltage of $V_g$ = 0 V, the current is considerably suppressed [Fig. 1(b)]. The on-off ratio for the data shown in Fig. 1(b), $I_d$ ($V_g$ = 40 V)/ $I_d$ ($V_g$ = –40 V), is approximately $10^4$. The $I_d$–$V_d$ curves [Fig. 1(a)] are nearly linear, indicating nearly ohmic characteristics of the contacts. In contrast, the device with Al contacts exhibits weak n-channel characteristics; $I_d$ slightly decreases with decreasing $V_g$ [Fig. 1(d)]. The threshold voltage, if it exists, should be much lower than –60 V. The $I_d$–$V_d$ curves exhibit nonlinear behavior, which is attributed to the Schottky barriers at the Al contacts. As shown in Fig. 2, the four-terminal $I$–$V$ characteristics were nearly linear, from which the sheet conductance $G_s$ was derived as a function of $V_g$. The result is compared with the sheet conductance of the device with Ti/Au contacts, as shown in Fig. 3. In the case of Ti/Au contacts, $G_s$ increases linearly above $V_g$ ~10 V and becomes virtually zero below $V_g$ = 0 V. In the case of Al contacts, $G_s$ increases slightly with increasing $V_g$, in accordance with the behavior of transfer characteristics. Furthermore, it is noteworthy that the value of $G_s$ for Al contacts is higher than that for the Ti/Au contacts for the range of $V_g$ used in this measurement. We note that $I_d$ of Al-contacted devices is often associated with large fluctuations as shown in Figs. 1(d), 2, and 3(b). The cause of the fluctuations has not been identified. We speculate that Al atoms that diffused into the MoS$_2$ layers, and are mobile, may be the possible cause of the fluctuations.

The transport properties of the Al-contacted devices drastically changed after Ti/Au layers were deposited onto the Al pads. The inset of Fig. 4(a) shows the photograph of the contact pads after the deposition of Ti (15 nm)/Au (75 nm) layers on the Al (100 nm) pads. The yellow Ti/Au layers are deposited on the Al pads, which appear in light yellow. For the overlapping parts, the boundaries of Ti/Au and Al pads nearly coincide with each other. The $I_d$–$V_g$ curves of the device before the deposition of Ti/Au layers and after the deposition of those on the Al pads are shown in Figs. 4(a) and 4(b), respectively. These data were obtained while the sample was in ambient air at room temperature. The large hysteresis shown in Fig. 4(b) is attributed to charge trapping associated with the molecules, mainly water, adsorbed on the surface of MoS$_2$.[23] Before the deposition of the Ti/Au layers, $V_g$-dependence of $I_d$ was very small, similar to that seen in Fig. 1(d). However, after the deposition of the Ti/Au layers on the Al pads, $V_g$-dependence of $I_d$ became very large, and the on/off ratio increased considerably, corresponding to the appearance of the off state at low values of $V_g$. In addition, $I_d$ decreased in the whole range of $V_g$. This result excludes the possibility that the diffusion of Al atoms into the MoS$_2$ layer is the origin of the suppression of the off state in the Al-contacted devices.

In the off state of the Ti/Au-contacted devices, the carrier is depleted when $V_g$ is below the threshold voltage. In contrast, the significant conductivity for the Al-contacted devices, that is, the absence of the off state, in the whole range of $V_g$ clearly indicates that there is a significant amount of carriers in MoS$_2$ in the whole range of $V_g$. We thus conclude that the observed difference between the transport properties (as shown in Figs. 1 and 3) in the devices with Ti/Au and Al contacts is attributed to the difference in the carrier densities in $MoS_2$; in the device with Al contacts, the carrier density is considerably higher than that with Ti/Au contacts. Because the modulation of the carrier density via gate voltage is a small fraction of the total carrier density in the device with Al contacts, $V_g$-dependence of $I_d$ is very small.

The cause of the difference in the carrier densities can be attributed to the difference in the work function of the metals. We assume that the work function of Ti (4.3 eV),[24] which acts as a sticking layer in the Ti/Au contacts, does not play a significant role in the carrier injection mechanism.[9] The literature values of the work functions of Au and Al are 5.40 eV and 4.54 eV, respectively.[24] The work function of $MoS_2$ is reported to be in the range from 4.5 to 5.2 eV.[9] The difference in work function means that charge transfer from $MoS_2$ to Au and from Al to $MoS_2$ may occur, causing an increased carrier density in $MoS_2$ with Al contacts. It should be noted that the actual values of the work functions may differ significantly from the literature values and that Fermi-level pinning is significant at the metal/$MoS_2$ interface. Therefore, an estimate of the Schottky barrier height based on the metal work functions is considered as approximate. However, we believe that the electron doping effect of the Al contacts is qualitatively explained in terms of the difference of the work functions of Al and Au. While we believe that the observed transport properties directly imply an electron doping effect by Al contacts, we plan to measure the carrier density by Hall effect measurement in the future to further prove this doping effect.

In clear contrast to the device with Al contacts, the device with Al/Ti/Au contacts (Al is contacting $MoS_2$, and Au is the top layer.) exhibited $I_d$–$V_d$ curves similar to that with Ti/Au contacts, as shown in Fig. 5. This implies that, for low values of $V_g$, the off state (nearly zero conductance) is realized. This observation is consistent with the result for a $WSe_2$ device with Al (10 nm)/Au (100 nm) contact.[8] In the cases of Al/Au and Al/Ti/Au contacts, the electron transfer from Al to Au occurs due to the difference in the work functions. This suppresses the electron doping effect of the Al contact on $MoS_2$. It is thus concluded that the electron doping effect of the Al contact is clearly observed only for a pure Al contact that is not in contact with high-work-function metals.

The $I_d$–$V_d$ curves for the Al contact are not linear but reflect the Schottky barrier between Al and $MoS_2$, as shown in Fig. 1(c). Similar characteristics attributable to Schottky barrier have also been reported in Ref. 8 for $WSe_2$ with Al/Au contacts with a more significant nonlinearity, which may be attributed to the decreased carrier density in $WSe_2$ due to the charge transfer from the Al to Au layer. The device with Al/Ti/Au contacts also exhibited similar $I_d$–$V_d$ curves (Fig. 5) to those of $WSe_2$ with Al/Au contacts. Based on the values of the work functions of Al and Au, one may expect that the Schottky barrier height is smaller for the Al contact than that for the Ti/Au contact,[6] which is inconsistent with the experimental results. Existence of high-resistance Schottky barrier for the Al contact can be attributed to the lack of d-orbitals in Al, which leads to a small overlap of electron orbitals with $MoS_2$.[8]

To demonstrate basic device performance that is compared with that reported in past literature, we estimated the field-effect mobility of our FET devices with Ti/Au contacts. The field-effect mobility is derived from $I_d$–$V_g$ curves as shown in Fig. 1(b), using the expression

$$\mu = \frac{dI_d}{dV_g} \cdot \frac{L}{WC_iV_d},$$

where $C_i$ = 13 nF/cm$^2$ is the capacitance between the channel and the back gate per unit area, and $W$ and $L$ are the channel width and length, respectively.[25] From the measurement of five devices (thickness ~12 nm), we obtained an average value of 13.2 cm$^2$V$^{-1}$s$^{-1}$ with a standard deviation of 2.9 cm$^2$V$^{-1}$s$^{-1}$. This value is considered to be the extrinsic mobility. The intrinsic mobility should be higher than this value because the above analysis excludes the effects of contact resistance and interface trap density.

**3.2 Temperature dependence**
Remarkable difference was observed in the temperature dependence of the $I_d$–$V_d$ curves for different contacts. Figures 6(a) and 6(c) show $I_d$–$V_d$ curves at various values of temperature for the devices with the Ti/Au and Al contacts, respectively. For the Ti/Au contact, the zero-bias two-terminal conductance $G_2$, that is, the slope of these curves at origin, decreases with decreasing temperature [Fig. 6(b)]. From the Arrehenius plot of $G_2$, the activation energy is derived as a function of $V_g$,[3,11,15,26,27] as shown in Fig. 7. The Schottky barrier height is then estimated to be about 150 meV. For the Al contact, the temperature dependence of the $I_d$–$V_d$ curve is considerably smaller than that for the Ti/Au contact. As shown in Fig. 6(d), the two-terminal conductance $G_2$ varies only by less than a factor of 2 in the temperature range of 13–290 K. It is remarkable that in the temperature range of 250–290 K, the temperature dependence between the two kinds of contacts is opposite; the conductance decreases with increasing temperature for the Al contact.

The weak temperature dependence for the Al contact indicates that the carrier injection mechanism at the contact is governed by the electron tunneling process, and not by any thermally activated processes such as thermionic emission and thermally assisted tunneling, both of which should be dominant in the electron transport for the Ti/Au contact. The assumed energy band diagrams corresponding to these electron injection mechanisms for the Ti/Au-contacted and Al-contacted MoS$_2$ FETs are shown in Figs. 8(a) and 8(b), respectively. In the case of $V_g > V_{th}$ for the Ti/Au contact, thermally assisted tunneling current is dominant, and is responsible for the linear $I_d$–$V_d$ characteristics,[11] where $V_{th}$ is the threshold gate voltage. In the case of $V_g < V_{th}$, corresponding to the off state, only thermionic emission current flows. In the case of the Al contact, the carrier density is significant with little dependence on $V_g$, and this is reflected by the nearly $V_g$-independent finite conductivity as shown in Figs. 1(c) and 1(d). This doping effect is shown by the lowering of the conduction band minimum [Fig. 8(b)]. Another important point in the case of the Al contact is the nonlinear $I_d$–$V_d$ characteristics [as shown in Fig. 1(c)], which are not significantly dependent on temperature. To explain these findings, we assume that the potential barrier at the Al/MoS$_2$ interface is higher but thinner, as compared to the case of the Ti/Au contact. Since the microscopic treatment of the electronic orbitals is required for full understanding, the discussion on the basis of these energy band diagrams should be considered as approximate and schematic. For the case of the Al contact, the energy band diagram shows little dependence on $V_g$. This is partly attributed to the screening effect of the carrier in the MoS$_2$ layers, which have thicknesses of approximately 15 nm in the samples we studied.

For some of the observed $I_d$–$V_d$ curves of the Al-contacted devices, plot of ln($I_d/V_d^2$) against $1/V_d$ implied a transition from direct tunneling to Fowler–Nordheim tunneling as shown in Fig. 9(a).[28-30] However, the weak temperature dependence, down to 13 K, as shown in Fig. 6(d), is considered as a direct evidence for electron tunneling, which does not involve thermal activation. This is because the analysis of the $\ln(I_d/V_d^2)$–$1/V_d$ plot is affected by the influence of sheet resistance on the $I_d$–$V_d$ curves and ambiguity in exponent in the formula for Fowler–Nordheim tunneling.[31] The Fowler–Nordheim plots for the Ti/Au-contacted device [Fig. 9(b)] is qualitatively different from those for the Al-contacted device. The decrease in $G_2$ with increasing temperature above 250 K for the Al contact can be attributed to the increase in sheet resistance due to enhanced phonon scattering. For tunneling that is not thermally assisted to be dominant, the potential barrier must be very thin. The thinning of the barrier is assumed to be related to the high carrier density induced by the electron doping effect of the Al contacts. Quantitative explanation of the observed $I_d$–$V_d$ curves would require simulations based on modeling of the potential barrier. Microscopic calculations such as density functional calculations would also be required for full understanding.[4,8,32]

## 4. Conclusions

We studied the transport characteristics of thin $MoS_2$ flakes contacted by Ti/Au and Al electrodes in a FET geometry. We assume that the Al contacts exhibited a significant electron doping effect, which was responsible for the absence of the off state in the devices studied within the range of back-gate voltages used in our experiment. While the temperature dependence of the two-terminal conductance for the Ti/Au contacts obeyed thermal activation, that for Al contacts was very weak. Furthermore, in the temperature range of 250–290 K, this temperature dependence for Al contacts was opposite to that for Ti/Au contacts. The weak temperature dependence implies that electron tunneling not accompanied by thermal activation is a dominant mechanism of carrier injection for Al contacts. It is known that ionic liquid gating using a very large electrical-double-layer capacitance is very effective in obtaining a very high carrier density in layered materials.[33,34] The doping effect induced by the Al contact may provide another useful method to induce a high carrier density in $MoS_2$ layers. These results highlighting the significance of the interfaces between contact metals and semiconductors are relevant to other two-dimensional semiconducting materials as well.

**Acknowledgment**

This work was supported by JSPS KAKENHI Grant Number 15K13497.

**Figures**

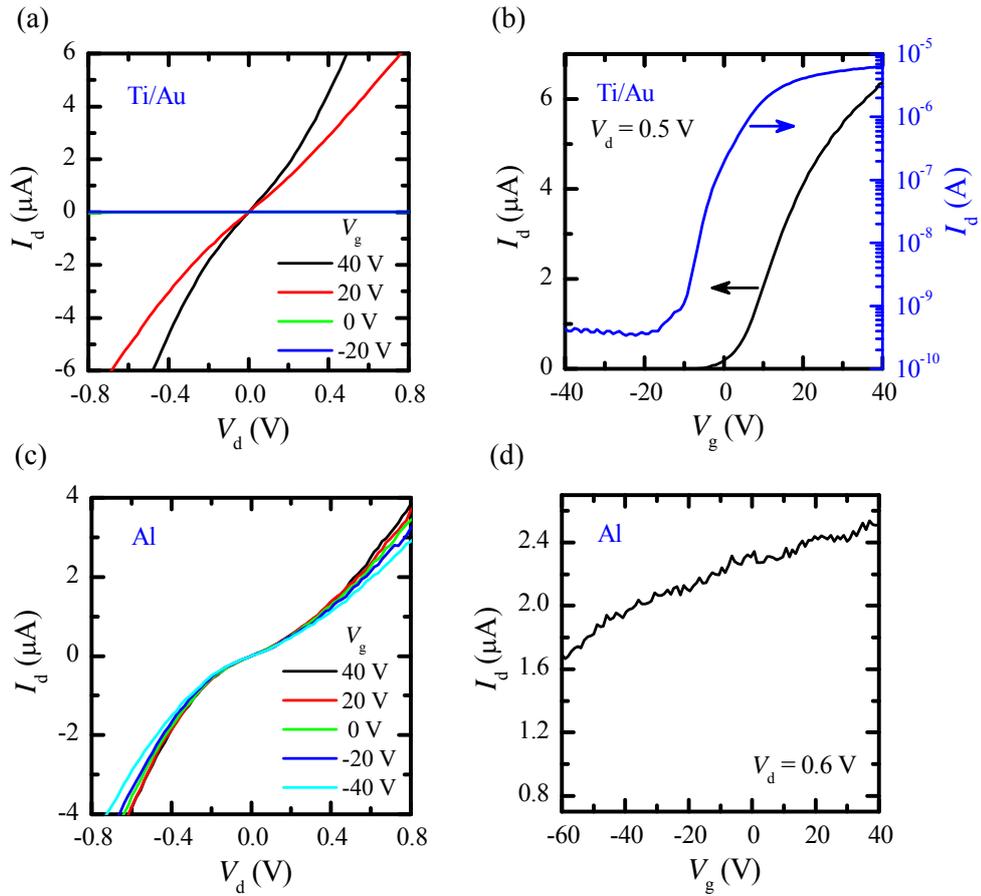

**Fig. 1.** (Color online) Output characteristics ($I_d$–$V_d$ curves) and transfer characteristics ($I_d$–$V_g$ curves) of MoS$_2$ field-effect transistors with Ti/Au contacts [(a) and (b)] and Al contacts [(c) and (d)] measured at room temperature. The $I_d$–$V_g$ curves (b) are shown on both linear and logarithmic scales. Device sizes (width/length) are 17 μm/5 μm for Ti/Au contacts, and 25 μm/24 μm for Al contacts. The thicknesses of the flakes are approximately 15 nm each.

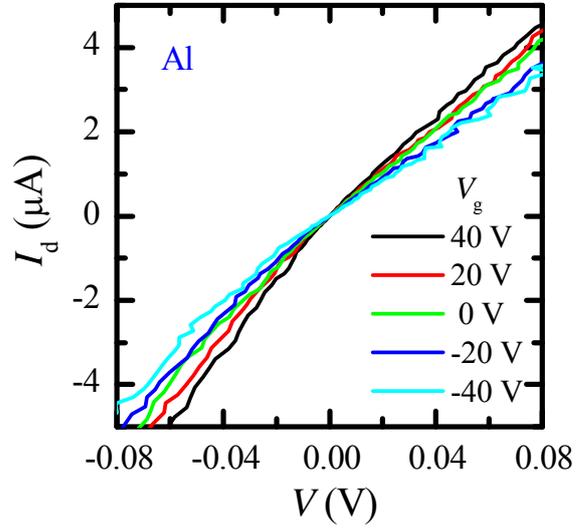

**Fig. 2.** (Color online) *I–V* characteristics of the device with Al contacts acquired by four-terminal measurement. The distance between the contacts for voltage measurement and the channel width are 16 μm and 27 μm, respectively. Nearly linear characteristics are shown.

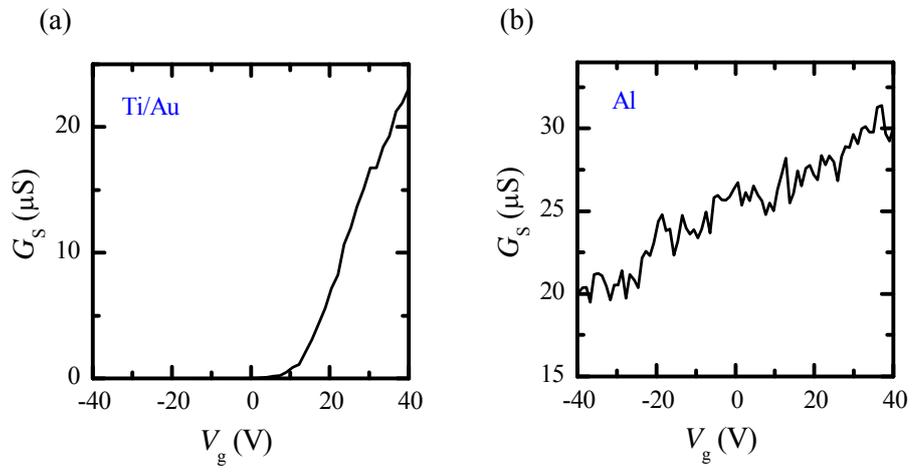

**Fig. 3.** (Color online) Sheet conductance $G_s$ of the MoS$_2$ flakes with Ti/Au contacts (a) and Al contacts (b) as a function of gate voltage $V_g$.

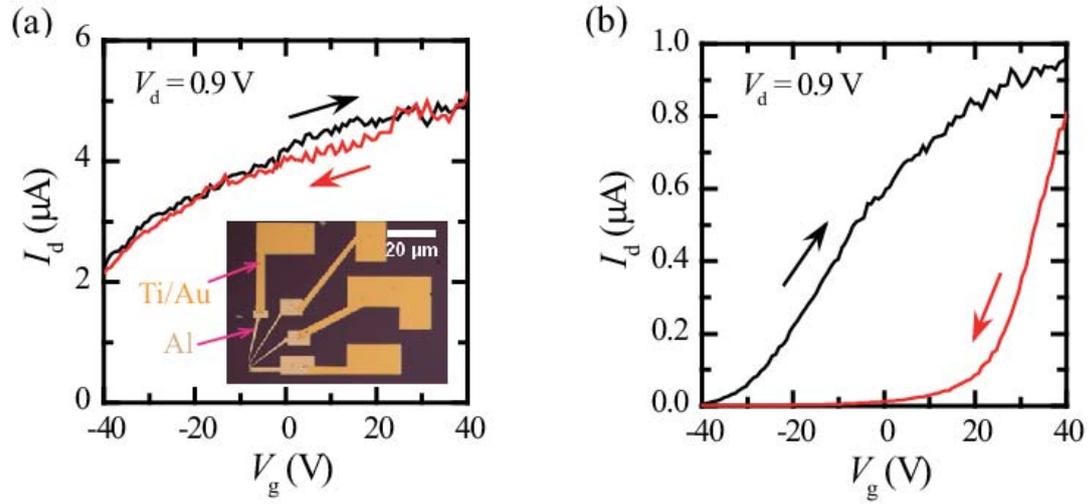

**Fig. 4.** (Color online) $I_d$–$V_g$ curves of a MoS$_2$ FET with Al contacts before the deposition of Ti (15 nm)/Au (75 nm) layers (a) and after the deposition of those on the Al (100 nm) pads (b). The inset of (a) shows the photograph of the contact pads after the deposition of Ti/Au layers, shown in dark yellow. The Al pads are shown in light yellow. The MoS$_2$ flake of 10-nm thickness is located in the lower left corner of the photograph. Two of the four contact pads shown in the photograph were used as the drain and source electrodes. These results were obtained in ambient air at room temperature. $V_d$ was fixed at 0.9 V. The sweep directions are indicated by the arrows. After the deposition of the Ti/Au layers, $V_g$-dependence of $I_d$ becomes very large, and the off state appears at low values of $V_g$. In addition, the values of $I_d$ shown in (b) are smaller than in (a) in the whole range of $V_g$, with significant hysteresis that is attributed to charge trapping. The device sizes (width/length) are 21 μm/4 μm.

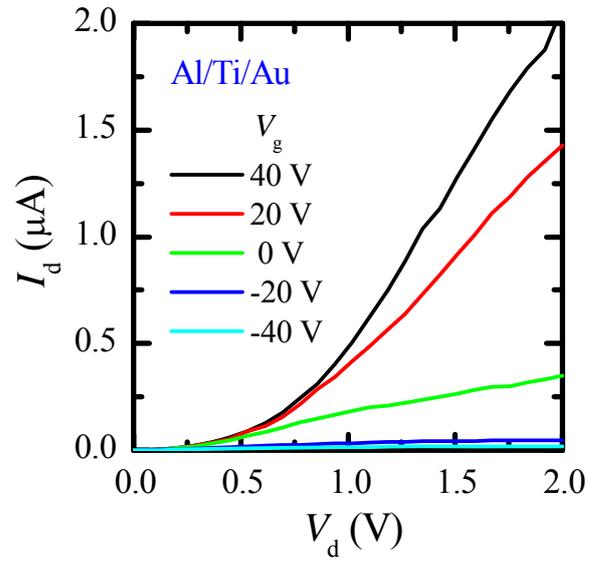

**Fig. 5.** (Color online) $I_d$–$V_d$ curves of the device with Al(3 nm)/Ti(17 nm)/Au(75 nm) contacts for $V_g$ = 40, 20, 0, –20, and –40 V. Conductance is nearly zero at $V_g$ = –40 V. Device sizes (width/length) are 17 μm/5 μm. These curves qualitatively agree with the $I_d$–$V_d$ curve of WSe$_2$ with Al/Au contacts.[8]

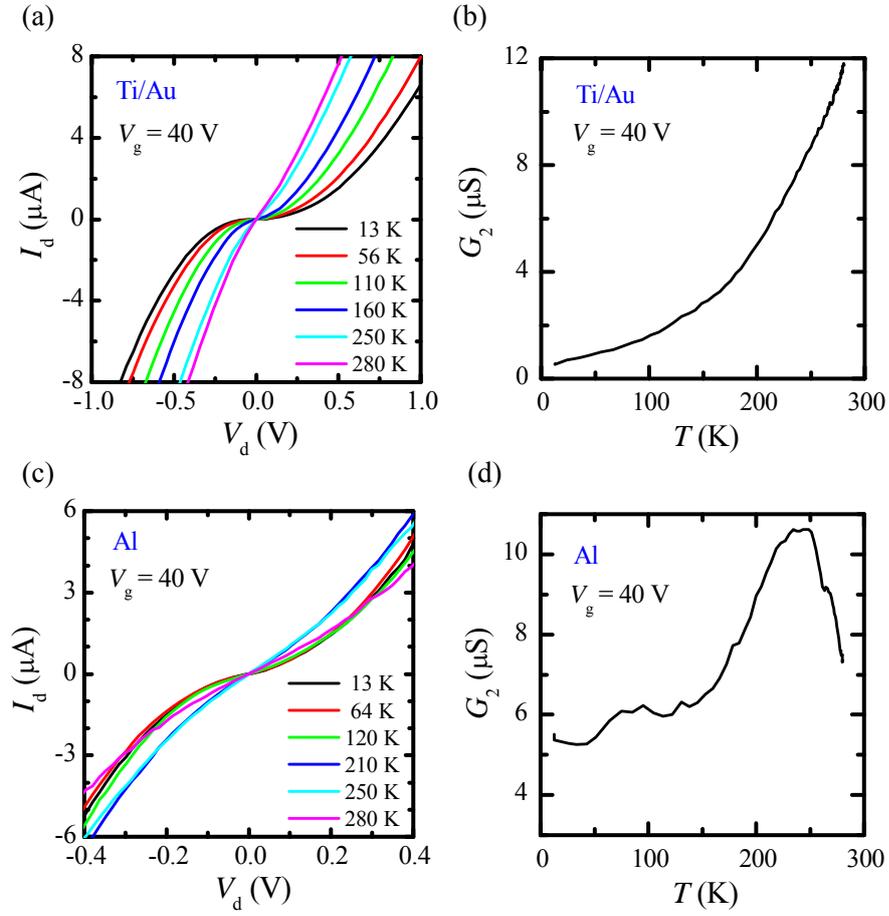

**Fig. 6.** (Color online) $I_d$–$V_d$ curves at various values of temperature and two-terminal conductance $G_2$ as a function of temperature for the devices with Ti/Au contacts [(a) and (b)] and Al contacts [(c) and (d)] for $V_g = 40$ V. Device sizes (width/length) are 17 μm/5 μm for Ti/Au contacts, and 27 μm/9 μm for Al contacts.

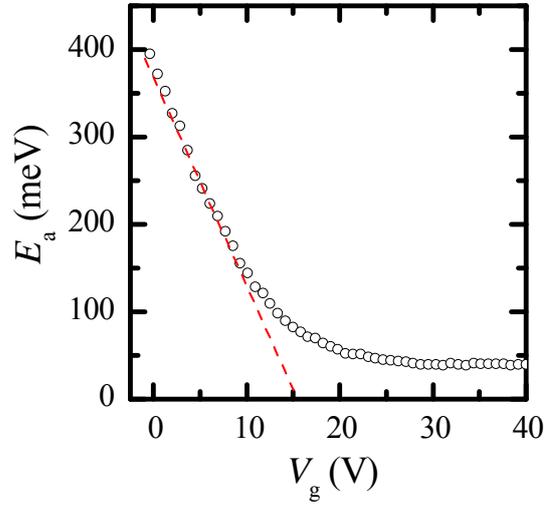

**Fig. 7.** (Color online) Activation energy $E_a$ of the two-terminal conductance extracted from the Arrehenius plot of $G_2$ for the device with Ti/Au contacts. The dashed line is a guide to the eyes. The true height of the Schottky barrier is estimated to be ~150 meV.[11] Above $V_g$ = 10 V, thermally assisted tunneling should be significant for carrier injection.

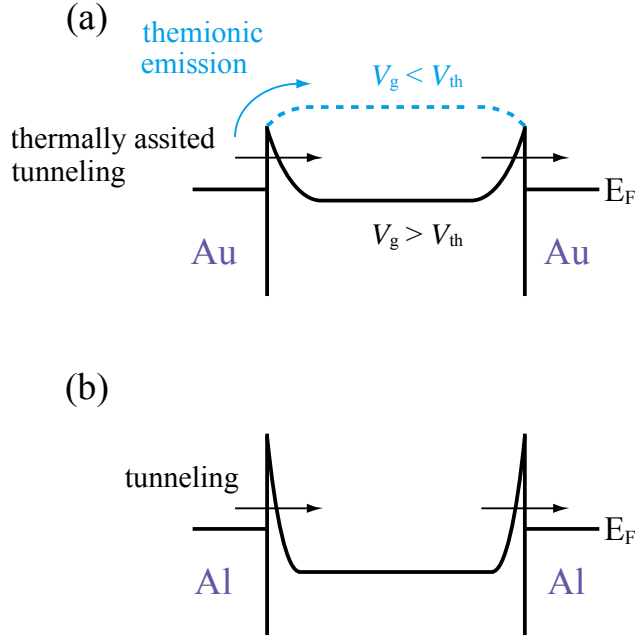

**Fig. 8.** (Color online) Energy band diagrams for Ti/Au-contacted (a) and Al-contacted (b) MoS$_2$ FETs. The curves corresponding to the levels of the valence band maximum and Ti sticking layers between MoS$_2$ and Au are omitted in this figure. The Fermi energy of the metal contact is denoted as $E_F$. Here, the drain-to-source voltage $V_d$ is assumed to be small. For the case of the Ti/Au contact (a), the band diagrams for $V_g > V_{th}$ and $V_g < V_{th}$ are shown by a black solid line and a blue dashed line, respectively, where $V_{th}$ is the threshold gate voltage. The off state corresponds to the condition $V_g < V_{th}$. While thermally assisted tunneling current is dominant for $V_g > V_{th}$, only thermionic emission current flows for $V_g < V_{th}$. For the case of the Al contact, the conduction band minimum is lowered, compared to (a), corresponding to the increased carrier density. The experimental results presented in this work imply that the carrier injection mechanism at an Al contact is dominated by tunneling that is not thermally activated.

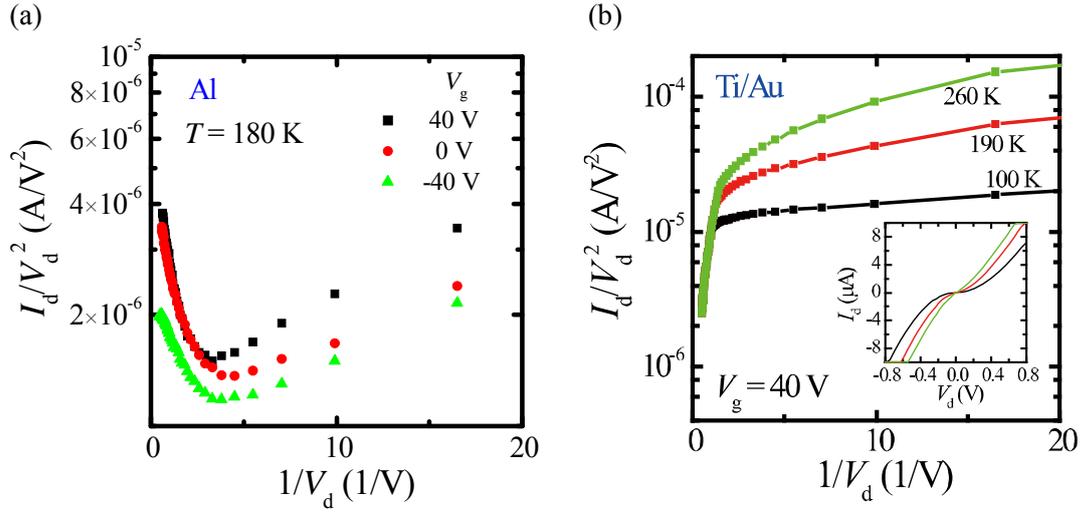

**Fig. 9.** (Color online) (a) Fowler–Nordheim plots corresponding to $I_d$–$V_d$ curves of the device with Al contacts at 180 K for $V_g$ = 40, 0, and –40 V. These plots may exhibit the transition from direct to Fowler–Nordheim tunneling. The dependence of the curves on $V_g$ is not significant and is in accordance with that of $I_d$–$V_d$ curves. (b) Fowler–Nordheim plots for the device with Ti/Au contacts at 260, 190, and 100 K for $V_g$ = 40 V. The inset shows the corresponding $I_d$–$V_d$ curves. These plots do not show the feature indicating Fowler–Nordheim tunneling.